# TF-UNet: Resolving Complex Speckles for Single-Shot Reconstruction of $512^2$-Matrix Images Using a Micron-Sized Optical Fiber

*Mingliang Xu$^{1,2}$†, Fangyuan Li$^{2,3}$†, Yuxin Leng$^{2,3}$, Ruxin Li$^{2,3}$ and Fei He$^{2,3}$\**

Mingliang Xu

[1]School of Physical Sciences, University of Science and Technology of China, Hefei 230026, Anhui, China

Mingliang Xu, Fangyuan Li, Yuxin Leng, Ruxin Li, Fei He

[2]State Key Laboratory of Ultra-intense Laser Science and Technology, Shanghai Institute of Optics and Fine Mechanics, Chinese Academy of Sciences, Shanghai 201800, China
E-mail: hefei@siom.ac.cn (F. H.)

Fangyuan Li, Yuxin Leng, Ruxin Li, Fei He

[3]Center of Materials Science and Optoelectronics Engineering, University of Chinese Academy of Sciences, Beijing 100049, China

Funding: National Natural Science Foundation of China (Grant No. 12388102), Zhangjiang Laboratory Youth Innovation Project (Grant No. S20240005), CAS Pioneer Hundred Talents Program and Shanghai Science and Technology Committee Program (Grant No. 23560750200).

**Abstract:** Tapered optical fibers (TFs), with diameters gradually reduced from hundreds of microns to the micron scale, offer key advantages over conventional flat optical fibers (FFs), including uniform illumination, efficient long-range signal collection, and minimal invasiveness for applications in high-sensitivity biosensing, optogenetics, and photodynamic therapy. However, high-fidelity, single-shot imaging through a single TF remains underexplored due to intermodal coupling from the tapering geometry, which distorts output speckle patterns and poses challenges for image reconstruction using existing deep learning methods. Here, we propose a physics-inspired TF-UNet architecture that augments skip connections with hierarchical grouped-MLP fusion to effectively capture non-local, cross-scale dependencies caused by intermodal coupling in TFs. We experimentally validate our method on both FFs and TFs, demonstrating that TF-UNet outperforms standard U-Net variants in structural and perceptual fidelity while maintaining competitive PSNR at quadratic complexity. Our study offers a promising approach for deep learning-based imaging through micron-sized, ultrafine optical fibers, enabling scanning-free single-shot reconstruction on a 512×512 reconstruction matrix, and further validating the framework on biologically meaningful neuronal and vascular datasets for physically interpretable characterization.

(† Mingliang Xu and Fangyuan Li contribute equally to the work.)

## 1. Introduction

Optical fiber components, including FFs, fiber bundles, and GRIN lenses, play a crucial role in advanced imaging and sensing by bridging precision photonics with biomedical engineering.[1-6] They address key limitations of traditional endoscopes, such as bulkiness and invasiveness, enabling cellular-level resolution, real-time deep-tissue imaging, and dynamic physiological monitoring.[7-9] However, challenges remain in maintaining signal fidelity, ensuring biocompatibility, and further miniaturization.[10-12] For example, FFs propagate multiple spatial modes, causing spatial dispersion and intermodal coupling.[13] FF bundles often face limitations in numerical aperture (NA) and spatial resolution due to pixelation and crosstalk between adjacent fibers. GRIN lens-based endoscopes typically employ shafts in the hundreds-of-micrometers range, have enabled deep-tissue imaging but are typically constrained by their bulky footprint and limited field of view, making long-term implantation challenging and potentially disruptive to surrounding biological tissues.[9] Even a recently reported "ultrathin" scanning endoscope has a shaft diameter of ~110 µm and relies on complex wavefront shaping or mechanical scanning, which introduce system-level complexity, sensitivity to motion artifacts, and increased power demands.[12] Currently, high-resolution, single-shot imaging using a single fiber with <10 μm core diameters remains a significant challenge.

More recently, tapered optical fibers (TFs), with ultra-thin tips have emerged as promising tools for bio-interfacing, enabling deep brain optogenetic modulation,[14] depth-resolved fiber photometry,[15] artifact-free fibertrodes,[16] and *in vivo* Raman spectroscopy for metabolic monitoring in mouse brains.[17] Compared with the flat optical fibers (FFs) with a highly localized light delivery, TFs gradually narrow in diameter along their length,[18] allowing more uniform illumination over larger areas. However, the axially varying geometry gives rise to intermodal coupling, and the axial evolution of the modal amplitudes is governed by the local propagation-constant mismatch $\Delta\beta$. As the waveguide dimensions change along the propagation direction, different modes experience varying effective refractive indices, causing their phase velocities to diverge. This results in a continuous power exchange between co-propagating modes, disrupting the stability of the signal transmission. Moreover, the spatial overlap between mode fields varies along the structure, further influencing the coupling dynamics. In fiber optics or integrated photonic devices, this phenomenon can degrade system performance by increasing modal dispersion and crosstalk, limiting bandwidth and transmission efficiency. While in fiber imaging reconstruction, compromises the effectiveness of inverse solvers based on multimode fibers—such as transmission matrix methods and deep learning approaches—by reducing their transferability across different configurations. To this end, geometry-aware inversion techniques capable of capturing non-local, cross-scale dependencies under tapered geometries are required.[3,19]

Meanwhile, AI-driven computational optics are emerging as a powerful solution to overcome these challenges. Various image reconstruction approaches are proposed for fiber applications in biomedical endoscopy to high-fidelity image transmission,[4,20] including methods based on transmission matrix theory,[21,22] real-time adaptive optics for dynamic aberration correction, and deep-learning-driven solutions.[23-35] Despite these advancements, conventional attention mechanisms fail to explicitly model space-variant mappings, which may compromise robustness under conditions of strong intermodal coupling. Although U-Net-based models [36] have demonstrated effectiveness in reconstructing images in turbid media, they exhibit limited capacity for non-local feature extraction [26,30,37] under complex optical fields or strong intermodal coupling. Although MLPs effectively model non-local dependencies via global spatial mixing, their O($n^2$) computational complexity poses limitations for high-resolution speckle image reconstructions.[38] Existing approaches face difficulties in capturing global contextual information owing to the nonlocal, space-variant nature of the forward mapping in TFs. Furthermore, most prior studies depend on datasets with over 10,000 samples or operate under simplified benchmark conditions, whereas practical deployment demands data-efficient models and that are capable of handling high-resolution inputs.

Overcoming these limitations requires either adaptive recalibration strategies, incorporation of spatial context into learning frameworks, or the development of physically informed models that explicitly account for the spatial variability of mode coupling and propagation dynamics within the fiber. In this study, we propose TF-UNet, which augments the skips with hierarchical grouped-MLP fusion to accommodate the space-variant mapping imposed by TFs. These physical priors are implemented via grouped-MLP fusion within skip pathways and an orthogonality regularizer. The grouped MLPs enable high-order, cross-channel, and spatial mappings at a channel-efficient quadratic computational cost, thus promoting mode disentanglement within skip connections. Simultaneously, the regularizer fosters decorrelated channel representations that align with physically meaningful mode separation. By training the model using intensity-only binary speckle–mask pairs from FFs and TFs, and evaluate their SSIM, MS-SSIM, LPIPS, [39,40] PSNR performance and Pearson correlation, we experimentally demonstrate that TF-UNet achieves superior structural and perceptual quality compared to U-Net baselines, while maintaining competitive PSNR levels with significantly reduced memory consumption. Our findings indicate that TF-UNet is a promising approach for high-fidelity image reconstruction using ultrathin, micron-sized fibers, supporting its potential for future robust *in vivo* applications.

## 2. Theory

As shown in **Figure 1(a)**, a TF is modeled as a step-index multimode fiber whose core radius $a(s)$ varies smoothly along the axial coordinate $s$. The local taper angle is defined as $\theta(s) = \arctan(da/ds)$. As the tapered end is typically used for excitation and collection, its conical sidewall will lead to geometry-related superposition of propagation modes. [14-16,18,30] This defines an $s$- and $\theta$-related waveguiding geometry in which the local V-number $V(s) = \frac{2\pi a(s)}{\lambda} \text{NA}$ changes along the taper. As *V(s)* varies, the set of supported linearly polarized (LP) modes and their propagation constants $\{\beta_m(s)\}$ evolve as well. Both the initial modal composition and propagation mode are spatially dependent, causing the distal speckle field to exhibit space-variant statistics, *i.e.*, the intensity distribution, correlation characteristics, and randomness of the speckle patterns change depending on where and how light is launched into the proximal end of the fiber. Consequently, calibration or learned inverse model—commonly employed for imaging or optical focusing through FFs—based on stationary assumptions in FF becomes fundamentally limited in its applicability. These models typically assume that the input-output relationship of the fiber remains consistent across different input conditions, but due to the space-variant nature of the speckle field, such assumptions hold true only within narrow axial and azimuthal ranges around a specific calibration point, and do not generalize reliably across different TF positions. [4,21-23]

Under weak guidance, the LP modes at each position s form an orthonormal basis that can be parallel transported along the taper. The geometric influence originates from the gradual variation in the cross-sectional profile: to leading order in the taper rate, the off-diagonal coupling scales as $\kappa_{mn}(s) \propto a'(s)/a(s)\, \Omega_{mn}(V(s))$, where $\Omega_{mn}(V(s))$ is an O(1) dimensionless overlap factor that enforces an azimuthal selection rule—coupling is strongest among modes with the same azimuthal index l, although small asymmetries may slightly relax this constraint. As the radius increases, the propagation constant separation between modal branches decreases following a power law $\Delta\beta_{mn}(s) \propto a(s)^{-2}$, which remains nearly independent of the numerical aperture (NA) away from cutoff. In an up-taper, these scaling behaviors oppose each other: $\kappa$ increases whereas $\Delta\beta$ decreases. Their ratio defines a local adiabatic parameter $\alpha(s)$, which grows with the strength of the geometric perturbation when measured against the intermodal propagation-constant spacing. A small $\alpha$ indicates adiabatic transport, while larger values identify non-adiabatic regions characterized by significant mode exchange.

Although non-adiabatic coupling is defined locally, its impact on image formation is cumulative: we treat the local coupling rate as a line density integrated along the propagation path to obtain a cumulative coupling metric over the taper. This metric grows toward the large-radius end—where a geometric factor related to $a(s)$ and $a'(s)$ is larger—and is further reinforced by a spectral factor that scales with the number of guided modes $M$. Near modal cutoff thresholds, small inter-modal propagation-constant spacings ($\Delta\beta \to 0$) increase the coupling coefficients $\kappa_{mn}$. A two-mode reduction then admits a Landau–Zener transition model, which requires a reduced taper rate precisely where new modes appear. Coupling to radiation modes is likewise promoted as the mode dispersion curve approaches the cladding light line, degrading reversibility/back-propagation fidelity.

For FFs with constant radius, $a'(s) \equiv 0$, the propagation operator is effectively diagonal in the local modal basis, suppressing intermodal power exchange and allowing only deterministic phase accumulation along each guided mode. By contrast, in TFs where $a'(s) \neq 0$, coupling between guided modes, as well as coupling from guided modes into radiative modes, is unavoidable and accumulates over the fiber length. Given the same terminal aperture, an imaging configuration in which light is injected through the tapered end and collected at the flat end is therefore intrinsically less stable and harder to invert than in an FF. The underlying reason is that the local coupling coefficient scales as $\kappa \sim a'/a$, whereas the intermodal propagation-constant mismatch scales as $\Delta\beta \sim 1/a^2$. **Figure S1** provides a quantitative comparison of the cumulative coupling metric and its axial density for a TF versus an FF with the same NA (see **Supporting Information**).

## 3. Experiment

### 3.1. Construction of TF speckle dataset

The dataset contains 13,440 speckle–mask pairs (6,720 for TF and 6,720 for FF) on a 512×512 reconstruction matrix. We randomly sampled 6,720 natural images from ImageNet [41] using stratified sampling to ensure class balance, and for each selected image we acquired one TF speckle measurement and one FF speckle measurement under their respective calibrated acquisition pipelines, and TF and FF observations share the same underlying targets. Due to the nonlinear mapping between light field intensity and image grayscale, and exponential growth of digital micromirror device (DMD) loading time for high-grayscale images, binary grayscale levels are used in the ImageNet experiments, and the dataset is split into training/validation/test subsets with an 8:1:1 ratio (5,376/672/672 pairs) for each fiber type. Preprocessing steps involved: (1) Centering and cropping to maximize square regions; (2) bicubic resizing to match the DMD's ROI; and (3) binary mask generation. Degraded speckle images were acquired using the DMD-to-fiber coupling setup: each preprocessed natural image was converted into a single-bit binary mask and displayed on a 2716×1600-pixel DMD (HDSLM54D67, ±17.5° mirror tilt, China) by intensity thresholding, which then illuminated the tapered end of the TF (NA = 0.39, active length = 2.5 mm, OptogeniX, Italy) using a 10× objective (NA = 0.25, WD = 10.6 mm, Olympus, Japan). The resulting output speckles were captured by a CMOS camera (STC-MBA1002POE, 3856×2764, Sentech, Japan) through a 20× objective (NA = 0.40, WD = 1.2 mm, Olympus, Japan) and a plano-convex lens ($f$ = 30 mm, Thorlabs, USA), with exposure time optimized (10 - 500 ms) to maximize signal strength. The dataset was split 8:1:1 into training, validation, and test sets, respectively.

**Figure 1(a)** illustrates the concept of image reconstruction using a TF: an ultrathin tapered optical fiber with tip diameter of 5 μm is used to collect and transmit the natural images, the generated speckle pattern at its output facet is used for reconstructing the original incident image using our proposed TF-UNet, which will be depicted later. **Figure 1(b)** shows the schematic and experimental system of a DMD-based setup illuminated by a continuous-wave 532 nm semiconductor laser (Lighthouse Photonics, USA). The laser output is collimated and expanded to generate a 9-mm-diameter beam, which then illuminates the DMD operating in off-axis blazed mode. A doublet lens ($f$ = 150 mm) relays the modulated beam to a 10× objective lens, which couples the image pattern into the tapered end of the TF. The numerical aperture of the objective is carefully selected to match that of the fiber (NA = 0.39), thus optimizing the trade-off between coupling efficiency and mode selectivity. After propagation through the TF, the output field forms a complex speckle pattern, which is collected by a 4f imaging system composed of a 20× objective (NA = 0.40) and a plano-convex lens (PCX, $f$ = 30 mm), and subsequently captured by a CMOS camera.

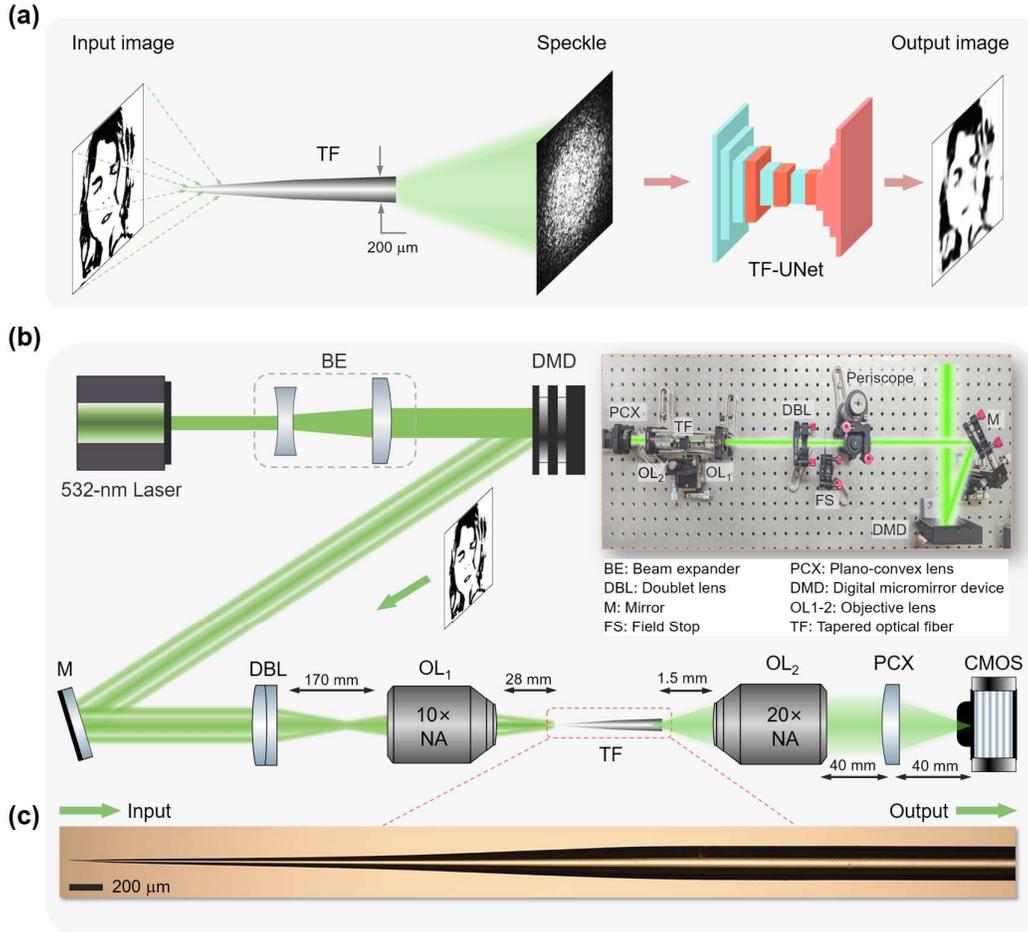

**Figure 1. Schematic illustration and experimental apparatus of learning-based image reconstruction using a TF.** **(a)** Schematic of the image dataset construction and the Tapered-Fiber UNet (TF-UNet) architecture. An input image is transmitted through a TF, generating a spatially variant speckle pattern, which is subsequently decoded by the TF-UNet to reconstruct the original image. **(b)** Optical layout for data acquisition: A 532-nm laser beam passes through an optional beam expander and is then directed to a digital micromirror device (DMD), which generates structured light patterns corresponding to natural images. The modulated light is relayed through an imaging system composed of an achromatic doublet lens (DBL) and a 10× objective ($OL_1$), projecting the image onto the tapered end of a TF. At the output end, a relay coupling optics system consisting of a 20× objective ($OL_2$) and a plano-convex lens (PCX) collect the resulting speckle pattern, which is captured by a CMOS camera. Inset shows the photograph of the experimental

setup. **(c)** Optical micrograph of a tapered optical fiber (NA = 0.39, active length = 2.5 mm) used for image reconstruction in (b).

## 3.2. Construction of TF-UNet architecture

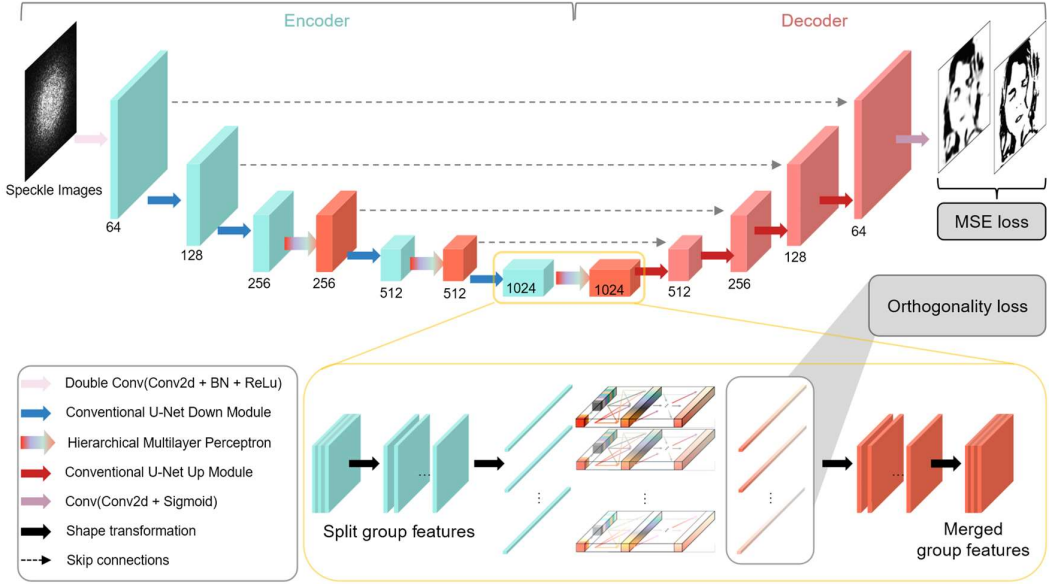

**Figure 2. TF-UNet architecture for decoding space-variant speckle fields.** An encoder–decoder backbone with four scales and skip connections (dashed arrows) processes the input speckle images. The numbers below indicate channel widths. At the bottleneck, a split–merge grouping module divides the channels into several groups along the channel dimension, applies a hierarchical MLP–based transformation to each group to enable geometry-aware, non-local mixing, and then concatenates and projects the outputs to form merged group features. An orthogonality regularizer is applied to ensure that the group bases are decorrelated, followed by a final $1\times1$ convolution with a sigmoid activation to produce the reconstruction.

We propose TF-UNet, a U-Net–based architecture tailored for speckle reconstruction through TFs, as illustrated in **Figure 2**. Inspired by the light propagation dynamics in TFs, where axially varying geometry induces space-variant input-output speckle mappings through intermodal coupling. TF-UNet enhances skip connections with a hierarchical feature-fusion block built upon grouped MLPs. This design enables non-local, high-order cross-scale interactions, allowing geometry-rich features from multiple encoder stages to exchange information prior to fusion in the decoder. Features are first refined using convolution layers and then processed by the grouped-MLP module to achieve cross-channel and cross-spatial fusion, effectively capturing geometry-related dependencies and promoting mode disentanglement under strong coupling. By incorporating the fusion within the skip pathways, TF-UNet targets high-frequency details and ensuring geometric consistency, while preserving the standard encoder–decoder backbone for reproducibility.

Specifically, we apply the hierarchical grouped-MLP fusion to the skip pathways using a staged grouping schedule across the four encoder scales (from shallow to deep: 0/1/2/4, where 0 indicates disabling the fusion at that scale). This schedule aligns the channel-mixing capacity with decreasing feature-map scale and can be adjusted according to task complexity and reconstruction-matrix size. When the grouping is set to 1/1/1/1, the block reduces to a standard ungrouped MLP. At each insertion point, the encoder feature $X \in \mathbb{R}^{B \times C \times H \times W}$ is divided into $G$ groups $\{X^{(g)}\}_{g=1}^{G}$, each containing $C_g$ channels. Each group $X(g)$ is then spatially flattened to length $N = H \cdot W$, resulting in $X^{(g)} \in \mathbb{R}^{B \times C_g \times N}$, followed by two shared nonlinear layers that operate along the spatial axis with each group

$$Z^{(g)} = \phi_2 \left( W_2^{(g)} \phi_1 \left( W_1^{(g)} X^{(g)} \right) \right) \qquad (1)$$

where each $\phi_k$ denotes Layer Normalization followed by ReLU, and $W_k^{(g)} \in \mathbb{R}^{N \times N}$ are spatial nonlinear mappings shared across the $C_g$ channels of group $g$. The outputs from all groups are reshaped back to $(H, W)$ and concatenated along the channel dimension to form the fused skip feature for the decoder. Per block, since each of the $G$ groups use one $N \times N$ spatial nonlinear shared across its channels, the dominant parameter and computational cost are $\Theta(GN^2)$ (two Linears: $2GN^2$), versus $\Theta(CN^2)$ for a per-channel spatial fully connected mapping. Thus, the grouped variant reduces the parameter count and computational load to approximately $(G/C) \times$ that of the per-channel baseline.

To promote channel-wise decorrelation that aligns with mode disentanglement, we introduce an orthogonality loss at the deepest stage of the encoder. Let $F \in \mathbb{R}^{C \times N}$ denote the mean-normalized feature matrix obtained by flattening the spatial dimensions, where $(N = H \cdot W)$, and arranging channels as rows. We compute the channel-wise Gram

$$G = \frac{1}{N}FF^\top \in \mathbb{R}^{C\times C} \quad (2)$$

and minimize the orthogonality defined as

$$L_{\text{ortho}} = \|G - I_C\|_F^2 \quad (3)$$

where $I_C$ is the $C \times C$ identity and $\|\cdot\|_F$ denotes the Frobenius norm. This loss encourages statistical independence across channels, thereby enforcing decorrelation.

The reconstruction loss is computed as the mean-squared error (MSE) between the predicted output $\hat{Y}$ and ground truth $Y$:

$$L_{\text{MSE}} = \frac{1}{n}\sum_{i=1}^{n}\|\hat{Y}_i - Y_i\|_2^2 \quad (4)$$

The overall objective function combines pixel-level accuracy with channel decorrelation:

$$L_{\text{total}} = L_{\text{MSE}} + \lambda L_{\text{ortho}} \quad (5)$$

where $\lambda$ controls the regularization strength.

We train the model end-to-end on a paired dataset of speckle patterns and target images acquired using the setup in **Figure 1(b)**. The network learns to map distorted speckles to reconstructions by minimizing $L_{total}$. Baselines, including U-Net variants and FF-oriented DL methods, follow the same training protocol. Training runs for 60 epochs on an NVIDIA A100 80 GB GPU with a batch size of 8, using SGD and ReduceLROnPlateau for learning-rate scheduling.

### 3.3. Evaluation of the indicators

In addition to PSNR ( ↑ ) and SSIM ( ↑ ), we report MS-SSIM ( ↑ ), LPIPS ( ↓ ), and Pearson correlation, CORR ( ↑ ) to jointly assess pixel accuracy, perceptual fidelity, and global statistical consistency. All metrics are computed on linearly scaled images (intensities range [0,1]) within a fixed, registered ROI. For decoding the speckle generated from a TF, we employ complementary metrics that reveal these trade-offs instead of masking them, as lower pixel error often comes at the cost of texture loss or structural drift across scales. MS-SSIM extends SSIM through a Gaussian pyramid decomposition, enabling the evaluation of both coarse structures and fine details across multiple scales. It combines the luminance at the coarsest level, $l_M$, with contrast–structure products at each level $j$ into a weighted geometric mean

$$\text{MS} - \text{SSIM}(x, \hat{x}) = l_M(x, \hat{x})^{\alpha_M} \prod_{j=1}^{M} mcs_j(x, \hat{x})^{\beta_j} \quad (6)$$

where the exponents are fixed and consistent across implementations. This multi-scale approach effectively detects cross-scale inconsistencies caused by intermodal coupling, which may differentially affect low-frequency layout and high-frequency edge fidelity along the transmission regions at the taper of TFs.

LPIPS measures perceptual dissimilarity in a fixed deep feature space rather than in pixel(RGB) space. Let $f_\ell(\cdot) \in \mathbb{R}^{H_\ell \times W_\ell \times C_\ell}$ denote the frozen feature map extracted from layer $\ell$ of a pretrained network $\phi$ (we use AlexNet with "lin" calibration). Inputs to $\phi$ are normalized to the range $[-1,1]$ to match the official LPIPS implementation. At each spatial location, features are L2-normalized across channels. The LPIPS distance is then computed by a channel-weighted sum of squared difference, averaged over spatial positions and aggregated across layers:

$$\text{LPIPS}(x, \hat{x}) = \sum_\ell \frac{1}{H_\ell W_\ell} \sum_{h,w} \left\| \mathbf{w}_\ell \odot \left( \hat{f}_\ell(x)_{h,w} - \hat{f}_\ell(\hat{x})_{h,w} \right) \right\|_2^2 \quad (7)$$

where $\mathbf{w}_\ell \in \mathbb{R}_{\geq 0}^{C_\ell}$ are the learned non-negative channel weights from the official calibration. Lower values indicate better similarity. Due to channel normalization and the use of deep semantic features, LPIPS is less sensitive to global luminance or contrast shifts compared to PSNR or single-scale SSIM, and better preserves judgments of texture and edge structure. However, it is not strictly invariant to photometric changes.

The reconstruction performance of TF-UNet was evaluated against five prominent U-Net variants—conventional U-Net,[36] recurrent residual-enhanced R2-UNet,[42] hybrid attention-driven TransUNet,[43] generative model-based VQVAE2[44] and MSMDFNet[26]—using a high-resolution dataset of 6,720 speckle-truth pairs under TF-specific physical constraints. MSMDFNet represents the current state-of-the-art for FF image reconstruction, whereas the others cover major deep learning paradigms in image restoration: CNN encoder–decoder architecture, recurrent residual network, Transformer-augmented models, and generative discrete-latent models, providing a balanced baseline. When applied to TF image reconstruction, these baseline models show notable limitations: R2-UNet's fixed-context recurrence captures short-range evolution but fails to model long-range, geometry-related correlations caused by taper-geometry–induced intermodal coupling and is sensitive to hyperparameter tuning. TransUNet offers global contextual awareness but becomes computationally prohibitive at 512×512 reconstruction matrix due to quadratic attention complexity and memory constraints, thus limiting its ability to resolve fine-scale diameter gradient correlation. In contrast, TF-UNet introduces an MLP-enhanced hierarchical fusion in the skip pathways, enabling learnable cross-scale and nonlocal interactions that are tailored to the geometric coupling characteristics of TFs. All quantitative results reported in this work are obtained from five independent training and evaluation runs under the same protocol using different random seeds and parameter initializations, and are summarized as mean ± standard deviation.

**3.4. Construction of neuronal and vascular validation datasets**

Neuronal and vascular datasets were additionally constructed to assess whether the proposed framework extends beyond binarized ImageNet masks under a feasible acquisition budget. For neuronal data, we used a deep-cortex calcium imaging dataset acquired using a freely moving miniature two-photon microscope (FHIRM-TPM 3.0).[45] From the recorded videos, 4,440 frames were extracted, uniformly rescaled, and converted into 512 × 512-pixel grayscale images quantified to 64 gray levels. We used 3,552 images for training, 444 for validation, and 444 frames from an independent video for testing, so that the test set does not share temporal continuity with the training stream. Under our current configuration, acquiring one neuronal speckle measurement (including pattern loading and CMOS exposure) requires approximately 1 minute per target image.

For vascular data, we used a rodent-brain vasculature dataset (MiniVess) obtained from two-photon microscopy, in which the original 3D stacks were binarized to highlight vessel structures.[46] During preprocessing, each 3D volume was partitioned along depth into multiple sub-volumes. For each sub-volume, slices were projected along the depth direction to form a 2D vascular image. Rotational augmentation was applied to expand the dataset to 4,000 binarized vascular images, which were split into 3,200/400/400 for training/validation/testing. For both validation datasets, target images were displayed on the DMD and corresponding speckle patterns were acquired through the same optical pipeline and registration procedure as in the ImageNet experiments, producing paired speckle–truth data for model training and evaluation under matched protocols.

## 4. Results

### 4.1. Image reconstruction using a TF

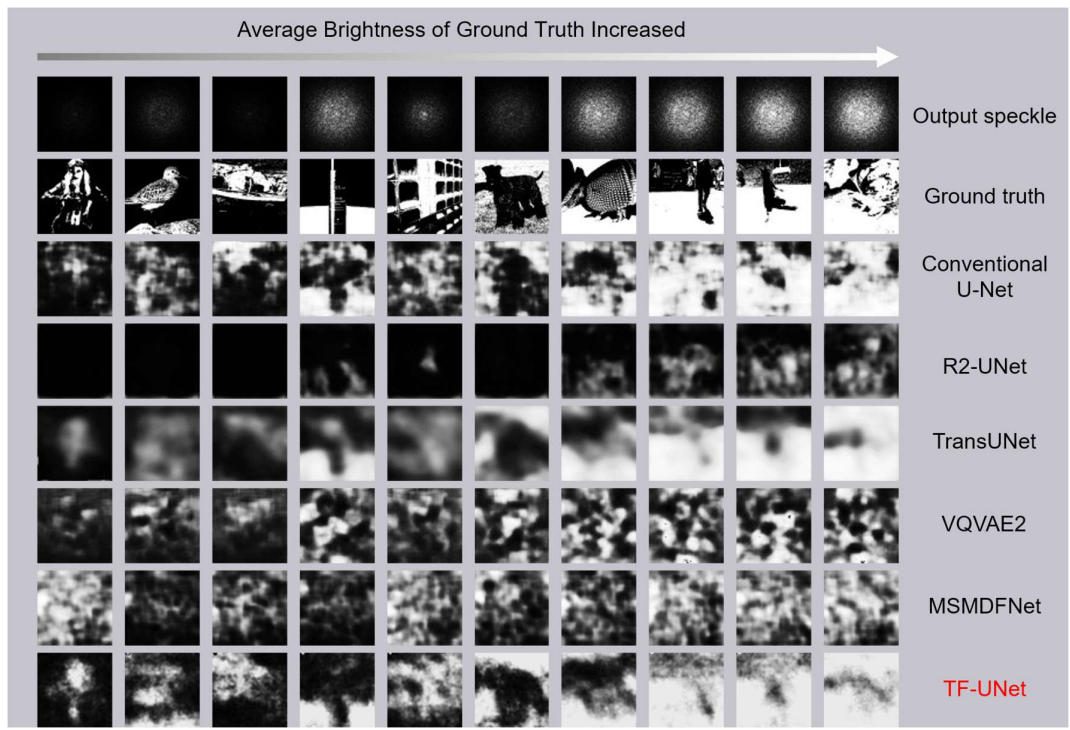

**Figure 3. Reconstruction test results of images from ImageNet and transmitted through the TF using various neural network approaches.** The test samples are arranged via deterministic sampling based on ascending mean grayscale values. Row 1: Original speckle patterns at the fiber output, showing severely degraded image quality. Row 2: Corresponding ground truth images. Other rows: Reconstructions from UNet, R2-UNet, TransUNet,

VQVAE2, and MSMDFNet, respectively. Last row: Reconstruction using the proposed TF-UNet.

As shown in **Figure 3** and **Table 1**, we benchmark TF-UNet against five representative architectures (UNet, R2-UNet, TransUNet, VQVAE2, and MSMDFNet) on the TF speckle–mask dataset using PSNR, SSIM, MS-SSIM, LPIPS, and Pearson correlation (CORR). Under the taper-induced, space-variant forward mapping, all methods operate in a strongly coupled regime, which compresses pixel-wise scores and makes perceptual/structural differences more salient. TF-UNet achieves the best SSIM and MS-SSIM, ties the best CORR with TransUNet (0.50), and remains competitive in PSNR (9.17 dB, second to TransUNet at 9.63 dB) and LPIPS (0.64, second to R2-UNet at 0.61). Among CNN-based baselines, UNet consistently outperforms MSMDFNet across all reported metrics, whereas VQVAE2 yields the lowest overall fidelity in this setting. The axial radius gradient in TFs introduces space-variant intermodal coupling (and leakage), which affects both global layout consistency and fine-scale texture recovery.

As a contextual comparison, we also evaluated the same reconstruction pipeline on a flat-cleaved fiber (FF), whose propagation is closer to a stationary input–output mapping and can therefore yield higher reconstruction scores under matched protocols. The FF results are provided in the **Supporting Information** (**Fig. S4** and **Table S4**).

**Table 1. Quantitative comparison of image reconstruction performance on TFs using different deep learning paradigms of neural networks.** The proposed TF-UNet achieves superior or competitive performance compared to baseline methods including UNet, R2-UNet, TransUNet, VQVAE2, and MSMDFNet. All results are averaged over five independent runs and with results reported as mean ± standard deviation. Best results are highlighted in red (highest for ↑, lowest for ↓), second-best in orange.

|          | PSNR ↑      | SSIM ↑        | MS-SSIM ↑     | LPIPS ↓       | CORR ↑        |
|----------|-------------|---------------|---------------|---------------|---------------|
| UNet     | 8.98±0.45   | 0.18±0.02     | 0.22±0.02     | 0.69±0.03     | **0.39±0.02** |
| R2-UNet  | 5.5±0.30    | **0.22±0.02** | 0.18±0.01     | **0.61±0.02** | 0.10±0.01     |
| TransUNet| **9.63±0.32** | **0.22±0.01** | **0.31±0.02** | 0.67±0.03   | **0.50±0.03** |
| VQVAE2 | 6.24±0.13 | 0.12±0.01 | 0.13±0.01 | 0.75±0.03 | 0.04±0.01 |
| MSMDFNet | 6.43±0.29 | 0.14±0.01 | 0.13±0.01 | 0.72±0.04 | 0.03±0.01 |
| **TF-UNet** | **9.17±0.27** | **0.24±0.02** | **0.32±0.02** | **0.64±0.05** | **0.50±0.03** |

Compared to the reconstruction using FFs, TF-UNet achieves a favorable balance in TF image reconstruction: it ranks first on SSIM/MS-SSIM, second in LPIPS (0.64 vs. 0.61 for R2-UNet), and ties the best CORR (0.50 with TransUNet), despite an overall decline in performance across metrics. These outcomes indicate TF-UNet's robustness to TF-induced non-uniform intermodal coupling, axial mode variation, and spatial heterogeneity, while reliably preserving structural details and perceptual quality.

**4.2. Biological validation on neuronal and vasculature imaging data**

The tapered optical fibers with a millimetre implantable active length have been widely applied in bio - interfacing with the animal brain.[14-18] This is largely due to their minimal invasiveness compared with conventional flat fibers, which thus allows for long - term stable deep brain interfacing in rodents. As shown in **Figure 4(a)**, the tapered diameter reduces mechanical mismatch between the fiber and neural tissue, minimizing chronic immune responses and glial scarring—common issues with rigid, bulky implants.[47-49] This allows the TF to remain close to neurons long-term, ensuring stable optical delivery and signal collection. The design also enables precise targeting of deep brain structures like the hippocampus, thalamus, and substantia nigra, which are critical for studying memory, sleep regulation, and Parkinson's disease, yet difficult to access with conventional optical imaging methods. We tentatively propose to use the TF for minimally invasive, implantable bioimaging reconstruction in the mouse deep brain, as shown in the proof-of-concept illustration in **Figure 4(a)**. To validate TF-UNet's applicability to biological samples, we applied the same TF acquisition and reconstruction pipeline on a 64-gray-level mouse brain calcium imaging and vasculature datasets, focusing on morphology recovery and its implication on downstream biological readouts.

As shown in **Figure 4(b),** TF-UNet produces markedly more coherent neuronal morphology than conventional UNet on calcium imaging reconstructions. The key factor for practical analysis is not the sharpness of isolated bright spots, but the network's ability to recover complete and continuous neuronal structures for stable region definitions. While UNet reconstructions are fragmented and morphologically incomplete, TF-UNet preserves more continuous and interpretable neuronal shapes with gray-level intensity variations. For physically interpretability, we measure the full width at half maximum (FWHM) measured on the annotated intensity profile, yielding ~15 μm for TF-UNet in a typical example. With well-recovered morphology, TF-UNet further enables detailed time series analysis of calcium events **(Figure 4(c-d))**. It maintains consistent spatial morphology across time, thus allowing stable ROI definition of individual neurons and reliable calcium trace extraction. Representative $\Delta F/F_0$ traces show that TF-UNet closely follows the ground-truth temporal fluctuations across multiple neurons (**Fig. 4(d)**). We quantify trace-level agreement using cosine similarity between reconstructed and ground-truth $\Delta F/F_0$ curves, achieving a Pearson correlation coefficient of 0.81 under the same protocol. These results indicate that TF-UNet reconstructions support calcium imaging through a TF, where reliable morphology recovery is essential for recovering intensity dynamics. We further test TF-UNet on a vasculature dataset **(Figure 4(e))**: from TF output speckles, TF-UNet reconstructs connected vascular networks with clear continuity and can resolve fine vessel structures at the tens-of-micrometers scale (representative examples annotated at 20 μm, 15 μm, and 15 μm, respectively). These test results indicate that the learned inverse mapping transfers to structured biological morphologies beyond the ImageNet-derived priors.

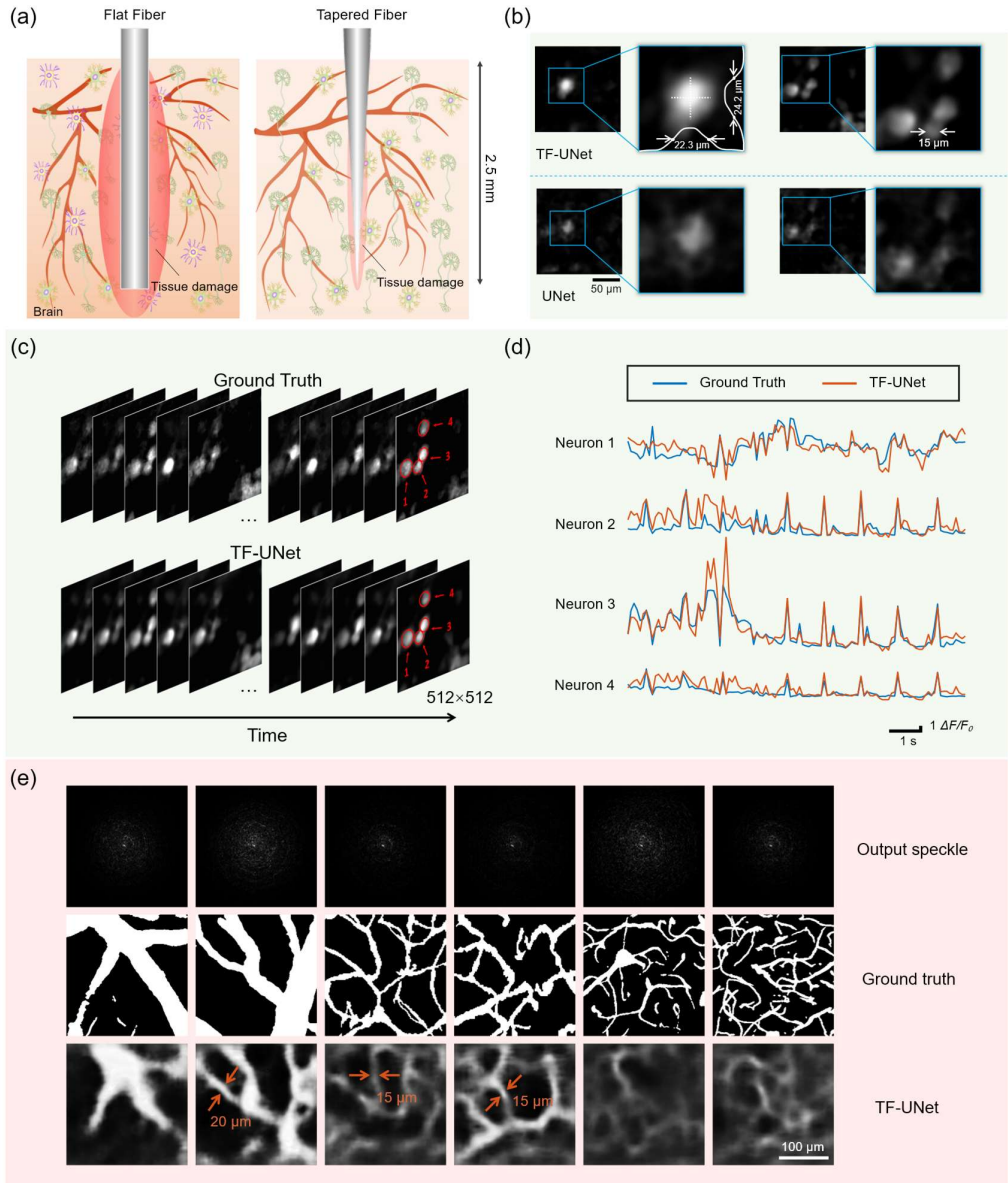

**Figure 4. Biological validation of TF-UNet on** rodent-brain **calcium imaging and vasculature data.** (**a**) Proof-of-concept comparison for mouse deep-brain imaging using a flat-cleaved fiber (FF) and a tapered fiber (TF). The tissue damage zone induced by the TF is much smaller than that induced by the FF, allowing for chronic stable deep-brain neuronal and vascular imaging. (**b**) Representative calcium imaging reconstructions from TF output speckles using TF-UNet and a conventional UNet. The insets highlight the close-up morphological continuity with a typical resolvable individual neuronal structure of approximately ~15 µm. (**c**) Example time sequence (512×512) showing ground-truth frames and the time series reconstructions by TF - UNet. The red circles and numbers highlight the representative ROIs of calcium events. (**d**) Extracted calcium trace ($\Delta F/F_0$) of the ground-truth (blue) and TF-UNet reconstructed (orange) neuronal activities as indicated in (**c**). (**e**) Validation of the TF-UNet modal on the vasculature dataset. TF output speckles (top), ground-truth vascular images

(middle), and TF-UNet reconstructions (bottom), with annotated examples highlight reconstruction of connected vessel structures at the capillary scale.

**Table 2. Quantitative comparison of image reconstruction performance on the Neuron dataset using different deep learning architectures. TF-UNet achieves the best SSIM and LPIPS, and remains competitive in PSNR/MS-SSIM/CORR compared with strong baselines including UNet, R2UNet, TransUNet, VQVAE2, and MSMDFNet.** All results are averaged over five independent runs and with variances reported. Best results are highlighted in red (highest for ↑, lowest for ↓), second-best in orange.

|  | PSNR ↑ | SSIM ↑ | MS-SSIM ↑ | LPIPS ↓ | CORR ↑ |
|---|---|---|---|---|---|
| Unet | 20.369 ± 0.272 | 0.661 ± 0.006 | 0.719 ± 0.009 | 0.358 ± 0.009 | 0.696 ± 0.023 |
| R2-UNet | 17.912 ± 0.197 | 0.443 ± 0.016 | 0.590 ± 0.009 | 0.445 ± 0.010 | 0.326 ± 0.022 |
| TransUNet | **24.651 ± 0.294** | *0.715 ± 0.004* | **0.809 ± 0.010** | *0.289 ± 0.005* | **0.865 ± 0.013** |
| VQVAE2 | 19.202 ± 0.210 | 0.539 ± 0.013 | 0.662 ± 0.010 | 0.394 ± 0.010 | 0.527 ± 0.023 |
| MSMDFNet | 20.299 ± 0.302 | 0.665 ± 0.006 | 0.769 ± 0.009 | 0.321 ± 0.010 | 0.784 ± 0.021 |
| Ours TF-UNet | *22.616 ± 0.220* | **0.765 ± 0.011** | *0.773 ± 0.009* | **0.285 ± 0.020** | *0.860 ± 0.015* |

Overall, the results in **Figure 4** support two conclusions: (i) the TF-UNet design is compatible with gray-level biological contrast where structural completeness (not only pixel-wise fidelity) is necessary, and (ii) under TF-induced space-variant coupling, TF-UNet can preserve morphology sufficiently to enable the recovery of time-dependent fluorescence dynamics, motivating its use in future TF-based functional bioimaging scenarios.

## 4.3. Performance, photon-coverage sensitivity, scale and reconstruction-matrix comparison

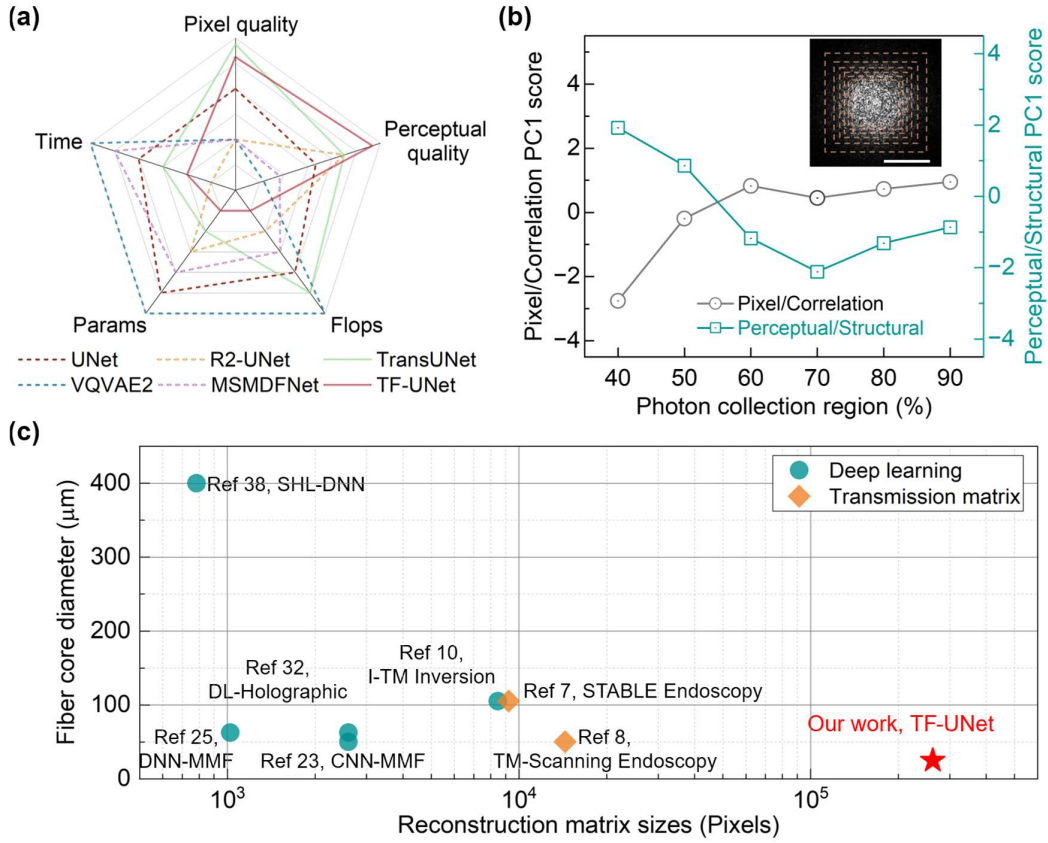

**Figure 5. Comparative evaluation of network performance, photon-coverage sensitivity, reconstruction matrix sizes and fiber-core footprints among existing methods and the proposed TF-UNet. (a)** Radar benchmark of UNet, R2-UNet, TransUNet, VQ-VAE2, MSMDFNet, and TF-UNet on pixel/correlation and perceptual/structural metrics, with efficiency (FLOPs, parameters, inference time). **(b)** Photon-coverage levels (40–90%) illustrated with nested apertures. PC1 from PCA captures metric variation with coverage. **(c)** Diagram showing the reconstruction matrix sizes and fiber core dimensions of different reconstruction approaches.

**Figure 5(a)** presents a radar chart summarizing model rankings across both accuracy and efficiency dimensions. The accuracy dimension is derived from the first principal component (PC1) of two metric families—Pixel and Perceptual scores—and is represented as a composite measure. Efficiency includes three aspects: multiply–accumulate operations (Flops), parameter count, and inference time. Scores for these metrics are inverted so that larger values correspond to higher efficiency. TF-UNet achieves the highest accuracy (Pixel = 5.25; Perceptual = 5.67) and ranks at or near the top in efficiency (inverted scores: 1.00, 1.00, 2.00), with 2762.88 GFlops, 172.66 M parameters, and 1.504 s inference time.

To better understand how photon collection regions affect reconstruction in TF imaging, we further analyzed pixel-wise and perceptual metrics, combining them in a principled manner. We categorized the metrics into two families: pixel-wise (PSNR and CORR) and perceptual (SSIM, MS-SSIM and LPIPS). Before aggregation, all metrics were z-score standardized across radii. LPIPS was transformed to $1-\text{LPIPS}$ so that larger values consistently indicate better quality. For each group, we applied PCA to the metric-by-radius matrix and used the first principal component (PC1) as a composite score **(Figure 5(b))**. As cumulative photon coverage increases from 40% to 90%, pixel PC1 increases steadily, whereas perceptual PC1 shows a clear minimum near $\approx 70\%$ coverage. This non-monotonicity arises from an annular marginal-contribution mechanism: expanding the sampling aperture from the bright core initially admits a mid-radius annulus with strong non-adiabatic, intermodal coupling and a highly space-variant forward mapping, causing local contrast and covariance mismatches that reduces perceptual quality ($\Delta Q(r) < 0$). Beyond this zone, irradiance and texture decay rapidly, making $\Delta Q(r) \to 0$, which dilutes earlier negative effects and leads to the rebound in perceptual PC1. In contrast, MSE–dominated measures like PSNR benefit from averaging over low-variance edge pixels, which explains the steady rise pixel PC1. These results motivate photon-efficient, spatially adaptive acquisition and learning—*e.g.*, coupling-aware attention, and loss re-weighting guided by PC1—to emphasize information-dense cores while regularizing the mid-radius, high-coupling zone. **Figure 5(c)** compares reconstruction matrix size (N×N pixels), against fiber core diameter across representative methods. Prior reports cluster at 28×28–120×120 pixels (784–14,400 pixels) with fiber/bundle core diameters of 50–400 μm, whereas TF-UNet with tapered optical fibers spans 20–200 μm and reaches 512×512 pixels (262,144 pixels) at micron-scale cores.

## 5. Discussion and Conclusion

We present TF-UNet for single-shot image reconstruction using a single micron-sized tapered optical fiber, in which complex speckle patterns are generated by axially varying geometry. Our approach demonstrates that integrating physics-inspired fusion directly into the skip connections of a U-Net architecture offers a powerful and effective strategy for addressing the challenges of space-variant intermodal coupling in computational imaging. By embedding physical models of light propagation into the network's information flow, the architecture inherently respects the underlying optical principles, enabling more accurate reconstructions while preserving computational efficiency. This approach not only improves quantitative performance metrics such as PSNR and SSIM in image restoration tasks but also enhances qualitative fidelity, particularly in regions with complex, geometry-related distortions that are difficult for traditional convolutional networks to model.[24,30,28,33,36,39]

Compared with conventional fiber-based techniques using coherent fiber bundles,[3,2,5,50] hundreds of micrometers-diameter GRIN lenses,[6,9,12,51] or ultrathin holographic/scanning endoscopes,[4,8] our methods occupied merely a distinct niche for widefield imaging by eliminating the need for active scanning or wavefront control — features that are critically important for implantable bioimaging and sensing applications. Relative to earlier fiber-imaging networks,[23,25,27,28,31,34] TF-UNet inserts physics-inspired fusion into the skip pathways to handle the space-variant, non-local transfer of tapered fibers, delivering higher pixel-wise and perceptual fidelity at 512×512 pixel while preserving deployable efficiency.

Despite these advantages, this study is subject to several limitations that constrain its current scope. First, supervision and evaluation are conducted at a single 532-nm wavelength with intensity-only measurements, which inherently neglect valuable phase information and multispectral cues that could enhance the model's understanding of the underlying physical properties of the samples. Phase contrast, for instance, often carries critical structural details—especially in transparent or weakly scattering specimens—that are invisible in intensity images alone. By relying solely on intensity data, the model may fail to capture subtle variations in refractive index or internal morphology, potentially limiting its accuracy in complex biological or material science applications. Moreover, the use of a 532 nm monochromatic light source limits spectral diversity, reducing contrast and specificity in identifying components within heterogeneous samples. Many biological molecules exhibit unique absorption or scattering properties across wavelengths, so multi-wavelength illumination can better distinguish structures with similar intensity but different spectral responses. Adding interferometric side channels, such as in digital holography or quantitative phase imaging, enables direct phase measurements, overcoming the absence of phase information.[17,11] This enriches the training signal and improves parameter identifiability, especially in complex inverse problems where different configurations yield similar intensity outputs. Extending the supervision framework to include multi-spectral inputs and outputs introduces additional physical diversity, boosting model performance. The current dataset draws primarily on natural-image statistics, which do not capture the distinctive microarchitecture and spectral signatures of biomedical imagery. Models trained on such priors often underperform on medical scans where precise localization and tissue classification are critical. Accordingly, evaluation should include *ex vivo* and *in vivo* data using fixed, spatially registered regions of interest to ensure cross-modality alignment, enable direct comparison with ground-truth annotations, improve reproducibility, and support longitudinal analyses across tissue types and imaging conditions.

Beyond metric gains, this work advances a transferable principle—co-designing neural modules with optical propagation—to address space-variant inverse problems in computational optics. Integrating physics with data-driven modeling yields systems that are more accurate and adaptive, reduces recalibration, and improves generalization. The approach is promising for deep-tissue fluorescence microscopy under scattering-limited conditions, enabling compact, hardware-efficient setups with high resolution and minimal illumination or labeling, and is likewise applicable to astronomy, remote sensing, and industrial inspection. [52-54]

**Acknowledgements**

This research was supported by the National Natural Science Foundation of China (Grant No. 12388102), Zhangjiang Laboratory Youth Innovation Project (Grant No. S20240005), CAS Pioneer Hundred Talents Program and Shanghai Science and Technology Committee Program (Grant No. 23560750200).

**Conflict of Interest**

The authors declare no conflicts of interest.